\documentclass{ws-ijmpd}
\usepackage[super,compress]{cite}

\usepackage{amsmath,amssymb,amsfonts}
\usepackage{graphicx}
\usepackage{color}

\def\noi{\noindent}

\def\nqq{\hspace*{-2em}}

\def\cm{\hspace*{1cm}}

\def\Jl#1#2{{\it #1}\ {\bf #2},\ }

\def\ApJ#1 {\Jl{Astroph. J.}{#1}}
\def\CQG#1 {\Jl{Class. Quantum Grav.}{#1}}
\def\DAN#1 {\Jl{Dokl. AN SSSR}{#1}}
\def\GC#1 {\Jl{Grav. Cosmol.}{#1}}
\def\GRG#1 {\Jl{Gen. Rel. Grav.}{#1}}
\def\JETF#1 {\Jl{Zh. Eksp. Teor. Fiz.}{#1}}
\def\JETP#1 {\Jl{Sov. Phys. JETP}{#1}}
\def\JHEP#1 {\Jl{JHEP}{#1}}
\def\JMP#1 {\Jl{J. Math. Phys.}{#1}}
\def\NPB#1 {\Jl{Nucl. Phys. B}{#1}}
\def\NP#1 {\Jl{Nucl. Phys.}{#1}}
\def\PLA#1 {\Jl{Phys. Lett. A}{#1}}
\def\PLB#1 {\Jl{Phys. Lett. B}{#1}}
\def\PRD#1 {\Jl{Phys. Rev. D}{#1}}
\def\PRL#1 {\Jl{Phys. Rev. Lett.}{#1}}

\def\lal{&&\nqq {}}
\def\eq{Eq.\,}
\def\eqs{Eqs.\,}
\def\beq{\begin{equation}}
\def\eeq{\end{equation}}
\def\bear{\begin{eqnarray}}
\def\bearr{\begin{eqnarray} \lal}
\def\ear{\end{eqnarray}}
\def\earn{\nonumber \end{eqnarray}}
\def\nn{\nonumber\\ {}}

\def\nnn{\nonumber\\ \lal }

\def\yy{\\[5pt] {}}
\def\yyy{\\[5pt] \lal }

\def\dst{\displaystyle}
\def\tst{\textstyle}
\def\fracd#1#2{{\dst\frac{#1}{#2}}}
\def\fract#1#2{{\tst\frac{#1}{#2}}}
\def\Half{{\fracd{1}{2}}}
\def\half{{\fract{1}{2}}}

\def\e{{\,\rm e}}
\def\d{\partial}

\def\arg{\mathop{\rm arg}\nolimits}

\def\sign{\mathop{\rm sign}\nolimits}

\def\const{{\rm const}}

\def\eqn#1{\eq\eqref{#1}}
\def\rf{\eqref}

\def\mn{_{\mu\nu}}
\def\MN{^{\mu\nu}}
\def\mN{_\mu^\nu}

\def\cE{{\cal E}}

\def\kappa{\varkappa}

\def\sph{spherically symmetric}
\def\ssph{static, spherically symmetric}

\def\bh{black hole}

\def\wh{wormhole}
\def\whs{wormholes}
\def\asflat{asymptotically flat} 

\begin{document}

\markboth{K.A. Bronnikov}
{Nonlinear electrodynamics, regular black holes and wormholes}


\title{NONLINEAR ELECTRODYNAMICS, REGULAR BLACK HOLES\\ AND WORMHOLES}
	
\author{K. A. Bronnikov}

\address
       {VNIIMS, Ozyornaya ul. 46, Moscow 119361, Russia;\\
        Institute of Gravitation and Cosmology, Peoples' Friendship University of Russia\\ 
          (RUDN University),  ul. Miklukho-Maklaya 6, Moscow 117198, Russia;\\
	National Research Nuclear University ``MEPhI'', Kashirskoe sh. 31, Moscow 115409, Russia;\\
	kb20@yandex.ru}
        
\maketitle

\begin{history}
\received{Day Month Year}
\revised{Day Month Year}
\end{history}

\begin{abstract}
  We consider spherically symmetric configurations in general relativity, supported by  
  nonlinear electromagnetic fields with gauge-invariant Lagrangians depending on the single 
  invariant $f = F_{\mu\nu} F^{\mu\nu}$. Static black hole (BH) and solitonic solutions 
  are briefly described, both with only an electric or magnetic charge and with both  
  nonzero charges (the dyonic ones). It is stressed that only pure magnetic solutions can be 
  completely nonsingular. For dyonic systems, apart from a general scheme of obtaining solutions
  in quadratures for an arbitrary Lagrangian function $L(f)$, an analytic solution is found for 
  the truncated Born-Infeld theory (depending on  the invariant $f$ only). 
  Furthermore, considering spherically symmetric metrics with two independent functions of time,
  we find a natural generalization of the class of wormholes found previously by Arellano and Lobo
  with a time-dependent conformal factor. Such wormholes are shown to be only possible for some
  particular choices of the function $L(f)$, having no Maxwell weak-field limit. 
\end{abstract}


\ccode{PACS numbers: 04.70.Bw, 04.20.Jb, 04.20.Gz}


\section{Introduction}

  Nonlinear electrodynamics (NED) appeared in the 1930s with Born and Infeld's effort to remove 
  the central singularity of a point charge and the related energy divergence by generalizing 
  Maxwell's theory \cite{B-Inf}, and the version of NED put forward by Heisenberg and Euler, 
  motivated by particle physics \cite{EuH}. It was further extended by Plebanski \cite{Pleb} 
  in the framework of special relativity, including an arbitrary function of the electromagnetic field
  invariants. 

  The interest in NED in modern studies is to a large extent motivated by the discovery that some 
  kinds of NED appear as limiting cases of certain models of string theory \cite{seiberg, tsey}. 
  It is also clear that the real electromagnetic field should lose its linearity at high energies due to
  interactions with other physical fields, and NED theories may be considered as a simplified 
  phenomenological description of these interactions. On the other hand, NED as a possible material 
  source of gravity is able to create various nonsingular geometries of interest, in particular, regular 
  black holes (BHs) and starlike or solitonlike configurations in the framework of general relativity 
  (GR) and alternative theories. 

  Among such models, the simplest are \sph\ ones, where the only possible kinds of electromagnetic 
  fields are radial electric and radial magnetic fields. Such solutions are widey discussed in the 
  literature, beginning probably with the paper by Pellicer and Torrence \cite{Pel-T} where a general 
  static solution was obtained for configurations with an electric field only. In Ref. \refcite{B-Shi},  
  a no-go theorem was proved showing that if NED is specified by a Lagrangian function $L(f)$
  (where  $f = F\mn F\MN$, and $F\mn$ is the Maxwell tensor), there is no such function $L(f)$ 
  having a Maxwell weak-field limit ($L\sim f$ as $f\to 0$) that a \ssph\ solution of GR with an 
  electric field has a regular center. This theorem was further extended to static dyonic configurations,
  with both electric and magnetic fields \cite{k-NED}, and it was further shown 
  \cite{k-NED, B-comment} that in numerous electric solutions describing configurations with 
  or without horizons (that is, BH or solitonic ones), having a regular center and a 
  Reissner-Nordstr\"om (RN) behavior at large radii $r$, there are different Lagrangian functions 
  $L(f)$ at large and small $r$: at large $r$ we have $L\sim f$ whereas at small $r$ the theory is
  strongly non-Maxwell ($f\to 0$ but $L_f \to \infty$, in agreement with the no-go theorem). 

  Meanwhile,\cite{k-NED} purely magnetic regular configurations, both BH and solitonic ones, are 
  possible and are readily obtained under the condition $L(f) \to L_\infty < \infty$ as $f  \to \infty$. 
  Electric models with the same regular metrics can be obtained from the magnetic ones using the 
  so-called FP duality \cite{k-NED} (not to be confused with the familiar electric-magnetic duality 
  in Maxwell's theory) that connects solutions with the same metric corresponding to {\it different\/} 
  NED theories. Unlike the magnetic solutions, the electric ones suffer serious problems 
  connected with multivaluedness of $L(f)$ and a singular behavior of the electromagnetic fields  
  on the branching surfaces \cite{k-NED}.

  Many results of interest were obtained since then, for a brief review see, e.g., Ref. \refcite{17-dyon} 
  and references therein. Among them let us point out a description of models with a kind of phase 
  transition allowing one to circumvent the above no-go theorem \cite{Bur1},  an extension of \ssph\
  NED solutions to GR with a nonzero cosmological constant $\Lambda$,\cite{Mat-09} 
  thermodynamic properties of regular NED BHs \cite{Bret-05, Kru-16, FW-16},
  cylindrically \cite{we-02} and axially \cite{Bam-13, Dym-15a} symmetric regular GR/NED  
  configurations and evolving wormhole models,\cite{Arel-06, Boe-07, Arel-09} 
  the stability properties of NED BHs,\cite{Mor-03, Bret-05s, Jin-14}
  and quantum effects in their fields.\cite{Mat-02, Mat-13}. One can also mention numerous 
  studies of special cases of electric and magnetic solutions, their potential observational properties 
  like gravitational lensing, particle motion and matter accretion in the fields of NED BHs, 
  their counterparts in scalar-tensor, $f(R)$ and multidimensional theories of gravity, inclusion of 
  dilaton-like interactions, non-Abelian fields, constructions with thin shells, etc., 
  but the corresponding list of references would be too long. The subject probably deserves a 
  comprehensive review.

  In this paper we discuss some recent progress concerning two subjects in the framework of 
  NED coupled to GR: \ssph\ dyonic solutions and \sph\ evolving \whs. The dyonic solutions 
  are inevitably singular at the center, as follows from the no-go theorem \cite{k-NED}, but they 
  are of interest as the first examples of this kind of solutions.\cite{17-dyon}. For completeness
  and comparison, pure electric and magnetic solutions are also briefly described.

  As to \whs\ as two-way tunnels or shortcuts between different universes or different, otherwise
  distant regions of the same universe, their possible existence and properties are widely discussed,
  see, e.g., Refs. \refcite{viss-book}--\refcite{WH-book} for reviews. 
  Wormholes are of interest not only as a perspective ``means of transportation'' but also as possible 
  time machines or accelerators \cite{thorne, BR-book}. Spherical symmetry is a natural simple 
  framework for \wh\ geometry, and most of known exact \wh\ solutions in GR and alternative theories 
  of gravity (e.g., Refs. \refcite{k73}--\refcite{BS-extra} and many others) 
  are \ssph. However, NED as a source of gravity in GR cannot support static \whs\ because 
  it does not provide the necessary violation of the Null Energy Condition (NEC). Only by considering
  evolving conformally static space-times it has been possible to obtain some examples of NED/GR 
  \wh\ solutions \cite{Arel-06}. In this paper, the approach developed in Refs.
  \refcite{Arel-06}--\refcite{Arel-09} is extended to a more general class of time-dependent metrics,
  containing two functions of time and somewhat similar to Kantowski-Sachs cosmologies.
  
  After presenting some general relations valid for both static and time-dependent NED/GR configurations 
  (Section 2), in Section 3 we briefly describe all three types of static solutions: magnetic, electric 
  and dyonic ones, following the previous papers, Refs. \refcite{k-NED} and \refcite{17-dyon}.
  In Section 4 we obtain and discuss a new class of nonstatic \sph\ \wh\ solutions, containing 
  those of Arellano and Lobo as a special case. Section 5 is a conclusion.

\section{Basic equations. FP duality}

  We start with the action 
\beq            \label{S}
	S = \Half \int \sqrt {-g} d^4 x [R - L (f)], 	
\eeq
  where $R$ is the Ricci scalar, $L(f)$ is an arbitrary functions, and 
  units are used with $c = 8\pi G =1$. The Einstein equations can be written, as usual, 
  in two equivalent forms
\bearr                                                                   \label{EE}
    G\mN \equiv R\mN - \half \delta\mN R = - T\mN, \qquad {\rm or}
\nnn \qquad
    R\mN = - (T\mN - \half \delta\mN T^\alpha_\alpha), 
\ear
  where $T\mN$ is the stress-energy tensor (SET), which in the theory \rf{S} is given by
  ($L_f \equiv dL/df$)
\beq             \label{SET1}
             T\mN = -2 L_f F_{\mu\alpha} F^{\nu\alpha} + \half \delta\mN L(f).
\eeq

  The metric is taken in the general \sph\ form 
\beq            \label{ds0} 
	ds^2 = A(x,t) dt^2 - B(x,t) dx^2 - r^2(x,t) d\Omega^2, \cm
						d\Omega^2 = d\theta^2+\sin^2 \theta d\phi^2.
\eeq
  The only nonzero components of $F\mn$ compatible with spherical symmetry are $F_{tr} =- F_{rt}$ 
  (a radial electric field) and $F_{\theta\phi} = - F_{\phi\theta}$ (a radial magnetic field). The 
  Maxwell-like equations $\nabla_\mu (L_f F\MN) = 0$ and the Bianchi identities 
  $\nabla_\mu {}^*F\MN = 0$ for the dual field $^* F\MN$ lead to
\beq              \label{F_mn}
                    r^2 \sqrt{AB} L_f F^{tr} = q_e, \cm    F_{\theta\phi} = q_m\sin\theta,
\eeq
  where $q_e$ and $q_m$ are the electric and magnetic charges, respectively. 

  Accordingly, the only nonzero SET components are 
\bearr        \label{SET2}
        T^t_t = T^r_r = \half L + f_e L_f, 
\nnn  
        T^\theta_\theta = T^\phi_\phi = \half  L - f_m L_f,
\ear
  and the invariant $f$ is the difference $f = f_m - f_e  = 2 ({\bf B}^2 - {\bf E}^2)$, where 
\bearr                   \label{ff}
	f_e = 2 F_{tr}F^{rt} = 2 {\bf E}^2 = \frac{2q_e^2}{L_f^2 r^4} \geq 0, 
\nnn  
	f_m = 2 F_{\theta\phi}F^{\theta\phi} = 2 {\bf B}^2 = \frac {2q_m^2}{r^4} \geq 0,
\ear
  ${\bf E}$ and ${\bf B}$ being the absolute values of the radial electric field strength and 
  magnetic induction, respectively, measured by an observer at rest in the reference frame under
  consideration. 

  The SET \rf{SET2} has the important properties $T^t_t = T^x_x$ and $T^t_x =0$; the latter 
  means the absence of radial energy flows, which is in turn related to the absence of 
  electromagnetic monopole radiation. Taken together, these two properties define a kind of matter
  sometimes called Dymnikova's vacuum \cite{dym-92, birk-16}, its evident vacuum-like property 
  is that its SET structure is insensitive to any transformations of the coordinates $x^0=t$ and 
  $x^1 =x$; in other words, all reference frames moving in the radial direction with any velocities 
  relative to each other are comoving to this kind of matter. Thus a \sph\ electromagnetic field 
  described by NED is a particular form of Dymnikova's vacuum.

  NED with a Largangian function $L(f)$ is known to admit a dual representation 
  obtained by a Legendre transformation \cite{Pel-T, sala-87, vag-14}: one defines the tensor 
  $P\mn = L_f F\mn$ with its invariant  $p = P\mn P\MN$ and considers the 
  Hamiltonian-like quantity 
\beq               \label{H}
	H(p) = 2f L_f  - L = - 2T^t_t 
\eeq
  as a function of $p$; then $H(p)$ can be used to specify the whole theory. One has then
\beq                                 \label{fp}
          L = 2p H_p - H, \qquad  L_f H_p = 1, \qquad  f = p H_p^2.
\eeq
  with $H_p \equiv dH/dp$. The SET in terms of $H$ and $P\mn$ then reads 
\beq                 \label{SET3}
	 T\mN =  -2H_p P_{\mu\alpha} P^{\nu\alpha} + \delta\mN (p H_p -\half H).
\eeq

  In a \sph\ space-time with the metric \rf{ds0}, \eqs \rf{F_mn} are rewritten in the P framework as
\beq                     \label{P_mn}
                     r^2 \sqrt{AB} P^{tr} = q_e, \cm    H_p P_{\theta\phi} = q_m\sin\theta.                       
\eeq
  We can also introduce the quantities $p_e$ and $p_m$ similar to \rf{ff}:  
\bearr                   \label{pp}
	p_e = 2 P_{tr}P^{rt} = \frac{2q_e^2}{r^4} \geq 0, \cm  
	p_m = 2 P_{\theta\phi}P^{\theta\phi} = \frac {2q_m^2}{H_p^2 r^4} \geq 0,
\ear
  so that $p = p_m - p_e$, and then the SET \rf{SET3} takes the form 
\bearr        \label{SET4}
        T^t_t = T^r_r = - \half H + p_m H_p, 
\nnn  
        T^\theta_\theta = T^\phi_\phi = -\half H - p_e L_f.
\ear
  Comparing \rf{SET2} and \rf{SET4}, one can see that they coincide up to the substitutions
\beq                \label{dual}
		\{F\mn,\ f,\ L(f)\} \ \  \longleftrightarrow \ \ \{^*P\mn,\ -p,\ -H(p)\},
\eeq
  where $^*P\mn$ is the Hodge dual of $P\mn$, so that $^*P_{\theta\phi}= P_{tx}$.
  The coincidence of the SETs means that the sets of metrics satisfying the Einstein equations
  \rf{EE} also coincide. It is the FP duality described in Ref. \refcite{k-NED}, which was 
  formulated there for \ssph\ systems and is extended here to nonstatic ones. It should be stressed 
  that this duality connects configurations with the same metric but in {it different\/} NED theories.
  An evident exception is the Maxwell theory, where $L = f = H = p$, and the FP duality turns
  into the conventional electric-magnetic duality. 

  If, however, one speaks of different formulations of the {\it same\/} theory, it turns 
  out \cite{k-NED, B-comment} that its $L$ and $H$ formulations are not always equivalent: 
  it is only the case if $f(p)$ is a monotonic function, see below.  

\section {Static systems}

  If the metric is static, so that the functions $A, B, r$ depend on $x$ only, for our system it 
  is reasonable to choose the ``Schwarzschild'' radial coordinate, $x=r$, so that $A=A(r)$ and 
  $B=B(r)$. Then, due to the equality $T^t_t = T^x_x$, we have from the Einstein equations 
  $R^t_t = R^x_x$, leading to $AB =1$ (with a proper choice of the time scale), so that 
\beq            \label{ds1} 
	ds^2 = A(r) dt^2 - \frac{dr^2}{A(r)} - r^2 d\Omega^2,
\eeq
  and the Einstein equation $G^t_t = -T^t_t$ then leads to 
\beq          \label{A}
	A(r) = 1 -\frac{2M(r)}{r}, \cm          M(r) = \frac 12 \int \cE(r) r^2 dr,  
\eeq
  where $\cE(r) \equiv T^t_t$ is the energy density, and $M(r)$ is called the mass function. It is 
  a general relation\cite{k-NED}, but it is only a part of a possible complete solution: the latter 
  requires a knowledge of $L(f)$ and both electric and magnetic fields.

\subsection{Pure magnetic and electric solutions} 

  Pure magnetic solutions ($q_e = 0$) are obtained from \eqs \rf{SET2} and \rf{A} quite easily.
  Indeed, if $L(f)$ is specified, then, since now $f = 2q_m/r^4$, the function $\cE(r) = L/2$ is known 
  from \rf{SET2}, and the metric function $A(r)$ is found by integration in \rf{A}. If, on the contrary, 
  $A(r)$ is known (or chosen at will), then $\cE(r) = L(f)/2$ is found from \rf{A}, and $L(f)$ is
  restored since $f = 2q_m/r^4$. A regular center requires $A(r) = 1 + O(r^2)$ at small $r$ 
  (and this leads to $L \to L_0 < \infty$ as $f\to\infty$ 
  \cite{k-NED}), asymptotic flatness requires $A(r) = 1 - 2m/r + o(1/r)$, where $m$ is the 
  Schwarzschild mass, and, if $\Lambda \ne 0$, asymptotically (A)dS solutions \cite{Mat-09} 
  are obtained by adding $ - \Lambda r^2/3$ to $A(r)$ in \rf{A}. It is an easy way to construct 
  regular magnetic BH and solitonic solutions, used by many authors, probably beginning with 
  Ref. \refcite{k-NED}.     

  A general feature of all such regular solutions is that $A\to 1$ as both $r\to 0$ and $r\to\infty$.
  Moreover, the mass term $-2m/r$ contributes negatively to $A(r)$ as long as $m > 0$
  whereas $q$ contributes positively as long as $L > 0$ (which is necessary for getting $A(0) =1$).
  Thus $A(r)$ in regular solutions inevitably has a minimum, and the value of $A$ at this minimum 
  --- hence the existence of horizons as zeros of $A(r)$ --- depends on a relationship between $m$ 
  and $q = q_m$. If the mass $m > 0$ is fixed, then at small $q$ the minimum of $A$ is negative 
  since the solution is close to Schwarzschild's in the whole space except small radii, $r < 2m$. 
  In this case each regular solution has two horizons, one of which is Schwarzschild-like, close to 
  $r=2m$, and the other exists since it is necessary to return to $A(r) > 0$ at small $r$ to reach $A=1$
  at $r=0$. At large $q$ the mass term becomes negligible at all radii except for the asymptotic region, 
  and a minimum of $A$ should become positive, leading to a solitonic solution. Therefore, some 
  value of $q$ should be critical, corresponding to a double zero of $A(r)$, hence a single extremal 
  horizon. This general picture is observed in all existing examples of regular \ssph\ NED-GR solutions. 
  We here illustrate it with the behavior of $A(r)$ in the example of Ref. \refcite{vag-14} in which
\beq                  \label{A-vag}
       A(r) = 1 - \frac{2m}{r} \biggl\{ 1 - \biggl[1+\biggl(\frac{2mr}{q^2}\biggr)^3\biggr]^{-1/3}\biggr\}.
\eeq   
  The behavior of $A(r)$ is depicted in Fig.\,1 for three values of $q/m$ leading to qualitatively different
  geometries. Their causal structures and Carter-Penrose diagrams are the same as those well known
  for RN space-times, with the important difference that now the lines $r=0$ correspond to a regular 
  center. 
\begin{figure}\centering
\includegraphics[width=6.5cm]{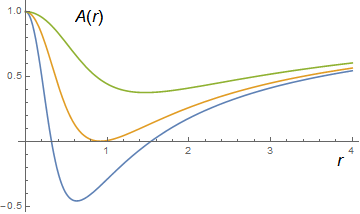}
\caption{$A(r)$ in BH and solitonic solutions given by \rf{A-vag} for 
	    $m=1$ and $q = 0.85,\ 1.0275,\ 1.3$  (bottom-up). } 
\end{figure}

  Pure electric solutions ($q_e\ne 0$, $q_m=0$) are obtained in a similar way by using the 
  Hamiltonian form of NED, see \eqs \rf{fp}--\rf{SET4}. We have simply $p= -2q_e^2/r^4$, 
  and specifying $H(p) = -2\cE(r)$, we directly find $M(r)$ and $A(r)$ from \rf{A}. On the 
  contrary, specifying $A(r)$, from \rf{A} we determine $\cE(r) = -H(p)/2$.  

  A regular center $r=0$ requires a finite limit of $H$ as $p \to \infty$. However, in any regular 
  \asflat\ (or (A)dS) solution $f = 0$ at both $r=0$ and $r=\infty$, so $f$ inevitably has at least 
  one maximum at some $p=p^*$, violating the monotonicity of $f(p)$, necessary for 
  equivalence of the $F\mn$ and $P\mn$ frameworks. As shown in Ref. \refcite{k-NED}, at an 
  extremum of $f(p)$ the Lagrangian function $L(f)$ suffers branching, its plot forming a 
  cusp, and different functions $L(f)$ correspond to  $p < p^*$ and $p > p^*$. Another kind of 
  branching occurs at extrema of $H(p)$, if any, and the number of Lagrangians $L(f)$ on the 
  way from infinity to the center equals the number of monotonicity ranges of $f(p)$ \cite{k-NED}. 
  An example of such behavior\cite{k-NED} is shown in Fig.\,2, it corresponds to the electric solution 
  with a regular center from Ref. \refcite{ABG-PLB}, in which  
\[
	H(p) = p \bigg/\!\cosh^2 \biggl[\frac{|q_e|^{3/2}}{2m}\Big(\frac {-p}{2}\Big)^{1/4}\biggr].	
\].
\begin{figure}[t]\centering
\includegraphics[width=7cm]{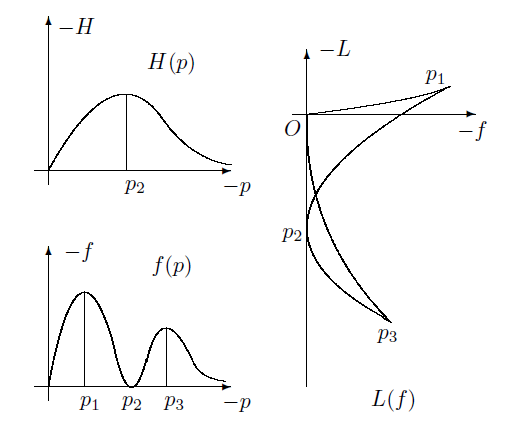}
\caption{Example of branching $L(f)$ in an electric solution}
\end{figure}

  The Hamiltonian framework might seem to be not worse than the Lagrangian one, even
  though the latter directly follows from the least action principle. However, it turns out\cite{k-NED} 
  that at $p=p^*$ the electromagnetic field is singular: the effective metric \cite{Nov-1, Nov-2} 
  in which NED photons move along null geodesics, is singular at extrema of $f(p)$, photons 
  are there infinitely blueshifted \cite{k-NED, Nov-2} and can create a curvature singularity by 
  back reaction on the metric. Thus any regular electric solution not only fails to correspond to 
  a fixed Lagrangian $L(f)$ but has other important undesired features.

  We conclude that each choice of $A(r)$ gives rise to both electric and magnetic solutions related 
  by FP duality. In these solutions the NED theories are different, and if $A(r)$ describes a 
  geometry with a regular center and a RN asymptotic at large $r$, only the magnetic solution
  is really regular.

\subsection{Dyonic configurations}

  Let us now assume that both $q_e$ and $q_m$ are nonzero. The difficulty in finding the solutions
  is that now $f(r)$ (or alternatively $p(r)$) is not known explicitly. Using, for definiteness, the 
  Lagrangian formulation of the theory, we have 
\beq                                     \label{f_r}
	 f(r) = \frac{2}{r^4} \biggl(q_m^2 - \frac{q_e^2}{L_f^2}\biggr). 
\eeq
  Comparing the expressions for $\cE(r)$ from \rf{SET2} and \rf{A}, we can write
  ($M' \equiv dM/dr$)
\beq    			      \label{M'}
           \Half L(f) + \frac{2q_e^2}{L_f r^4} = \frac{2M'(r)}{r^2} = \cE(r). 
\eeq

  Let us specify the theory by choosing $L(f)$. Then \eqn{f_r} can be considered as either 
  (A) an equation (in general, transcendental) for the function $f(r)$ or (B) an expression of $r$ 
  as a function of $f$. 

  In case A, if $f(r)$ can be found explicitly, integration of \eqn{M'} (equivalent to \rf{A}) 
  gives the metric function $A(r)$, and this completes the solution.

  Scheme B leads to a solution in quadratures in terms of $f$ which can be chosen as a new 
  radial coordinate. Indeed, assuming that $L(f)$ and $r(f)$ are known and monotonic, 
  so that $L_f \ne 0$ and $r_f \ne 0$, \eqn{M'} can be rewritten as 
\beq                                   \label{M_f}
               M_f = \frac{r^2 r_f}{2}\biggl[\frac{L}{2} + \frac{q_e^2}{L_f r^4}\biggr]
\eeq  
  (the subscript $f$ denotes $d/df$). Since the r.h.s. of \rf{M_f} is known, 
  it is straightforward to find $A(r)$ and to pass on to the coordinate $f$ in the metric.
  This gives us {\it a general scheme of finding dyonic solutions} under the above conditions.
  
  Consider two examples using scheme A.
  The {\bf first example} is used to verify the method: the Maxwell theory, $L=f$, $L_f=1$. 
  Then from \eqn{M'} we obtain $2M' = (q_e^2+q_m^2)/r^2$, whence 
  $2M(r) = 2m - (q_e^2+q_m^2)/r$ and 
\beq                    \label{RN}
                     A(r) = 1 - \frac{2m}{r} + \frac{q_e^2+q_m^2}{r^2}, \ \ \ m=\const,
\eeq 
  that is, the dyonic RN solution, as should be the case. 

  In the {\bf second example} we assume that \eqn{f_r} is linear with respect to $f$, which
  unambiguously leads to the truncated Born-Infeld Lagrangian,
\beq                  \label{BI}
	L(f) = b^2 \big(-1 + \sqrt{1+ 2 f/b^2}\big), \ \ \ b = \const
\eeq
  (the full Born-Infeld Lagrangian would also contain the invariant $(^*F\mn F\MN)^2$).
      
  Indeed, \eqn{f_r} is linear in $f$ only if $L_f^{-2} = c_1 f + c_2$, 
  $c_{1,2} = \const$. Integration gives $L = L_0 + (2/c_1)\sqrt{c_1f + c_2}$. For a Maxwell 
  behavior ($L {\approx} f$) at small $f$ we put $c_2 =1$, $L_0=-2/c_1$. Denoting $2/c_1 = b^2$, 
  we arrive at \eqn{BI}, and
\bearr              \label{f-BI}
            f(r) = \frac{2b^2 (q_m^2 - q_e^2)}{4q_e^2 + b^2 r^4},   \cm
	   \cE(r) = -\frac{b^2}{2} + \biggl(\frac{b^2}{2} + \frac{2q_e^2}{r^4}\biggr) S(r),
\ear
  where 	
\[		
             S(r) = \sqrt{1 + \frac {2f}{b^2}} = \sqrt{\frac{4q_m^2 + b^2 r^4}{4q_e^2 + b^2 r^4}}.
\]  
   The simplest solution corresponds to the special case of a self-dual electromagnetic field, 
   $q_e^2 = q_m^2$, whence $S(r) = 1$, $f=0$, and $L_f =1$ as in the Maxwell theory. This leads to
   $\cE = 2q_e^2/r^4$ and the dyonic RN metric with $A(r)$ given by \rf{RN}.

   In the general case $q_e^2 \ne q_m^2$, $M(r)$ and $A(r)$ are expressed in terms of 
   the Appell hypergeometric function $F_1$: 
\bear
                 2M(r) &=& 2m + \frac{12 q_e^2 S(r) +  b^2 r^4 (1 + 3 S(r))}{6 r} 
\nn 
            &&\  + \frac{b^2 r^3}{42 |q_e q_m|}
		 \biggl [14 (q_e^2 + q_m^2) 
       F_1\biggl(\frac 34, \Half, \Half, \frac 74, -\frac{b^2 r^4}{4 q_e^2}, -\frac{b^2 r^4}{4 q_m^2}\biggr) 
\nn 
             && \cm   + 3 b^2 r^4 
            F_1 \biggl(\frac 74, \Half, \Half, \frac{11}{4}, -\frac{b^2 r^4}{4 q_e^2}, 
                                       -\frac{b^2 r^4}{4q_m^2}\biggr)\biggr],
\ear
   where $m = \const$, and $A(r) = 1 - 2M(r)/r$. 

   Some features of the solution should be noticed. As expected, the center $r=0$ is singular in 
   accord with the above no-go theorem. At large $r$ the quantities
   $f_e$. $f_m$ and the energy density $\cE$ decay as $r^{-4}$, and the solution is \asflat\ 
   and approximately RN. At small $r$, $f(r)$ tends to a finite limit while 
   $\cE \approx 2 |q_e q_m|/r^4$. In a pure electric solution we also have $\cE \sim r^{-4}$, 
  while in a pure  magnetic one $\cE  \sim r^{-2}$. Thus all solutions are singular at the center 
   $r=0$, but the in the pure magnetic solution the singularity (existing since the function \rf{BI}
   does not tend to a finite limit as $f\to \infty$) is milder.  

\section{Dynamic wormholes}

  Apart from BH and solitonic configurations with a regular center, there can exist regular objects 
  having no center at all, namely, wormholes and some classes of regular BHs, including the so-called 
  black universes (BHs in which beyond the event horizon, instead of a singularity, there is an 
  expanding universe), see, e.g., Refs. \refcite{bu1}--\refcite{bu3}. In the static case all of them 
  require violation of the Null Energy Condition that states $T^t_t - T^x_x \geq 0$. Since for the 
  SET \rf{SET2} such a difference is zero, this condition (though marginally) is observed by NED, 
  so static \whs\ are manifestly impossible in the theory \rf{S}. However, as shown by Arellano and 
  Lobo \cite{Arel-06}, \wh\ solutions can be obtained if we consider evolving space-times. 
  We will try to extend their finding, considering a more general class of time-dependent metrics.  

\subsection {Equations}

  In Ref. \refcite{Arel-06} and the subsequent discussion \cite{Boe-07, Arel-09}, the metric was
  chosen to be static times a time-dependent conformal factor. Let us assume a more general metric,
\beq                        \label{ds2}
	 ds^2 = \e^{2\gamma(x) + 2\nu(t)}dt^2 - \e^{2\alpha(x) + 2\eta(t)} dx^2
			- \e^{2\beta(x)+2\omega(t)} d\Omega^2,
\eeq
  and NED as the matter source of gravity as given in \rf{S}. The function $\nu(t)$ can be  
  absorbed by redefinition of $t$ and $\alpha(x)$ by redefinition of $x$, with the result 
\beq                        \label{ds3}
	 ds^2 = \e^{2\gamma(x)}dt^2 - \e^{2\eta(t)}dx^2 - \e^{2\beta(x)+2\omega(t)} d\Omega^2
\eeq
  that substantially simplifies the Einstein equations. For this metric the nonzero 
  components of the Ricci tensor are
\bearr           \label{R00}
	R^t_t = \e^{-2\gamma}[2\ddot\omega + \ddot\eta + 2 \dot\omega{}^2 + \dot\eta{}^2]
			- \e^{-2\eta}[\gamma'' + \gamma'(2\beta'+\gamma')],
\yyy              \label{R11} 
         R^x_x = \e^{-2\gamma}[\ddot\eta + \dot\eta(2\dot\omega+\dot\eta)]
			- \e^{-2\eta}[2\beta''+\gamma'' + 2\beta'{}^2  +\gamma'{}^2],
\yyy
   R^\theta_\theta = R^\phi_\phi =  \e^{-2\gamma}[\ddot\omega + \dot\omega(2\dot\omega+\dot\eta)] 
			- \e^{-2\eta}[\beta''+ \beta'(2\beta'+\gamma')],
\yyy
	R_{tx}  = 2\dot\omega \beta' - 2 \dot\omega \gamma' - 2 \dot\eta \beta'.
\ear

  The electromagnetic field equations lead to the same relations 
  \rf{F_mn} and \rf{SET2} as in the static metric, where now $r = \e^{\beta(x)+ \omega(t)}$, 
  preserving its geometric meaning of the spherical radius. As before, the $(xt)$ component of the 
  SET is zero, and the corresponding Einstein equation $R_{xt} =0$ takes the form
\beq                   \label{01}
                       \dot\omega (\beta' - \gamma') = \beta' \dot \eta,
\eeq
  where dots denote $\d/d t$ and primes $\d/\d x$. Assuming $\beta' \ne 0$,
  $\dot\omega \ne 0$ and dividing \eqn{01} by $\beta' \dot\omega$, we separate the 
  variables and, without loss of generality, obtain
\beq                 \label{egam}
	\eta(t) = b\omega(t), \cm \gamma(x) = (1-b) \beta(x), 
\eeq 
  where $b$ is the separation constant. Next, since $T^t_t = T^x_x$, we have the equation 
  $R^t_t = R^x_x$, where, excluding $\gamma$ and $\eta$ according to \rf{egam}, 
  we again separate the variables, obtaining
\beq                     \label{bom} 
            \e^{2b\omega} [\ddot\omega + (1-b) \dot\omega{}^2]
                = - \e^{2(1-b)\beta} [\beta'' + b \beta'{}^2] = -k^2,
\eeq
  with the separation constant $k^2 {>\,} 0$ assumed to be positive since we are seeking solutions 
  with a minimum ($\beta' {=}0, \beta'' {>\,} 0$) of the function $\beta(x)$, describing a throat in 
  3D spatial sections $t=\const$.

  Thus $\beta(x)$ and $\omega(t)$ are determined by the equations
\bearr                     \label{b-eq} 
	                 \beta'' + b \beta'{}^2 =  k^2 \e^{2(b-1)\beta},   
\yyy                        \label{o-eq}
            		 \ddot\omega + (1-b) \dot\omega{}^2 = -k^2 \e^{-2b\omega},
\ear
  whose first integrals are easily found. Specifically, for $b \ne 1/2$ we have 
\beq                      \label{int1}
		\beta'{}^2 = C_1 \e^{-2b\beta} + \frac{k^2}{2b-1} \e^{2(b-1)\beta},\cm
		\dot\omega{}^2 = C_2 \e^{2(b-1)\omega} + \frac{k^2}{2b-1} \e^{-2b\omega}, 
\eeq
  and for $b =1/2$  
\beq  	           	  \label{int2}
		\beta'{}^2 = (C_3 + 2k^2 \beta) \e^{-\beta}, \cm
		\dot\omega{}^2 = (C_4 - 2k^2 \omega) \e^{-\omega},
\eeq
  where $C_{1-4}$ are integration constants. 

\subsection {Geometry}

  Solutions to \eqs \rf{int1} and \rf{int2} completely determine the metric under the ansatz \rf{ds2} 
  or \rf{ds3} for {\bf any} kind of matter whose SET satisfies the conditions  $T^t_t = T^x_x$ and
  $T_{tx} =0$. As already mentioned, these conditions define the so-called Dymnikova 
  vacuum \cite{dym-92, birk-16}, for which any reference frame in radial motion is comoving.

  In all cases under consideration, by construction, any 3D section $t = \const$ of space-time contains 
  a throat defined as a minimum of $\beta(x)$ and hence of $r= \e^{\beta(x)+\omega(t)}$
  at any given time instant, whereas the global features of these space-times depend on the 
  constants involved. 

  Some immediate observations on the time dependence of $\omega$.follow directly from \eqn{o-eq}.
  Thus, a regular minimum of $\e^\omega$ (that is, a bounce in the time evolution of $r(x,t)$) is 
  impossible since \eqn{o-eq} leads to $\ddot\omega < 0$ at points where $\dot \omega =0$ and 
  $\e^\omega >0$. Next, a finite limit of $\e^\omega$ as $t \to \pm\infty$ is also impossible because 
  at such a minimum the l.h.s. would be zero while the r.h.s. is finite. So the only possible way of 
  regular evolution of $\e^\omega$ is to begin or end with $\e^\omega \to 0$ at infinite proper time, 
  which is not completely excluded.  

  Further integration of \eqs \rf{int1} for $b\ne 1/2$ leads in the general case to the 
  hypergeometric function ${}_2 F_1$: 
\bearr                     \label {sol1}
                 \pm t =  \bigg| \frac{2b-1}{k^2}\bigg|^{1/2} \frac{\e^{(2-b)\omega}}{2-b}
	{}_2 F_1 \biggl(\Half, 1-\frac b2, 2- \frac b2, \frac{2b-1}{k^2}C_2 \e^{2\omega}\biggr),
\nnn
                 \pm x =  \bigg| \frac{2b-1}{k^2}\bigg|^{1/2} \frac{\e^{(1+b)\beta}}{1+b}
	{}_2 F_1 \biggl(\Half, \frac{1+b}2, \frac{3+b}2, \frac{2b-1}{k^2}C_1 \e^{2\beta}\biggr),
\ear
  while integration of \eq \rf{int2} for $b = 1/2$ involves the error function Erf:
\bearr                     \label {sol2}
		\pm t = \frac {\sqrt{\pi}}{k} \e^{-C_4/(4k^2)}
			\biggl[ -1 + {\rm Erf}\,\bigg(\frac{\sqrt{C_4 - 2k^2 \omega}}{2k}\bigg)\biggr],
\nnn
		\pm x = \frac {\sqrt{\pi}}{k} \e^{C_3/(4k^2)}
			\biggl[ -1 + {\rm Erf}\,\bigg(\frac{\sqrt{C_3 + 2k^2 \omega}}{2k}\bigg)\biggr].
\ear
  Both solutions \rf{sol1} and \rf{sol2} are rather hard for further investigation, let us therefore 
  restrict ourselves to some simple special cases. 

\medskip\noi
{\bf Example 1: $b =1$.} In this special case we obtain $\eta(t) \equiv \omega(t)$ and thus restore 
  the isotropic nature of space-time evolution considered in Refs. \refcite{Arel-06}--\refcite{Arel-09}.
  The metric takes the form
\beq                               \label{ds4}
           ds^2  = dt^2 - \e^{2\omega(t)} \big( dx^2 + \e^{2\beta(x)} d\Omega^2 \big).
\eeq 
  Assuming $b=1$ in \eqs \rf{int1} and \rf{int2}, we obtain the following solutioon 
  under a proper choice of the zero points of the coordinates $x$ and $t$:
\bearr                                               \label{isotr}
                  \e^\beta(x) = r_0 \cosh (kx), \quad\ r_0 = \const;
\qquad
                 \e^\omega = \left\{\begin{array}{ll} 
				        (k/h) \sinh (ht), \ & \ h>0,\yy
					    kt,                     & \ h=0,\yy
					(k/h) \sin (ht),    & \ h <0,   \end{array}\right.
\ear
  where we have denoted $C_2 = h^2 \sign h$. It can be directly verified that this solution coincides 
  with the one considered in Refs. \refcite{Arel-06}--\refcite{Arel-09}, but in another,
  more preferable parametrization:  in \rf{isotr} the coordinate $t$ is proper time, and $x$ 
  is proportional to the proper distance at any fixed time $t$.\footnote
	{Our coordinates $(t, x)$ and the coordinates $(t=t_{\rm AL}, r=r_{\rm AL})$ chosen  
         by Arellano and Lobo in Ref. \refcite{Arel-06} are connected by the relations
         $dt_{\rm AL} = \e^{-\omega(t)} dt$, $r_{\rm AL} = \e^{\beta(x)}$.
         The coordinate $r = r_{\rm AL}$ of Refs. \refcite{Arel-06}--\refcite{Arel-09} should
	not be confused with the quantity $r(x,t)$ used here, having the geometric meaning of
         the radius of spheres $\theta=\const, \phi=\const$. }  
  Thus the spatial sections are evidently of \wh\ nature but not \asflat\ (the latter would require
  $r \propto |x|$ at large $|x|$ whereas here $r \propto \e^{k|x|}$), while the time evolution is 
  different for different $h$: at $h\geq 0$ it is semi-infinite, and at $h < 0$ it occupies a finite 
  period between two zeros of $\sin (ht)$. The zeros of $\e^\omega(t)$ correspond to 
  big-bang type singularities.

\medskip\noi
{\bf Example 2: $b=0$.} In this case there is no expansion or contraction in the radial direction,
  and the coordinate $x$ is the true radial length in the reference frame used. The metric has the form
\beq 		\label{ds5}
		ds^2 = e^{2\beta(x)}dt^2 - dx^2 - e^{2\omega(t) + 2\beta(x)}d\Omega^2.
\eeq 
   Integration of \eqs \rf{int1} gives:
\beq 		\label{ds5a}
		\e^\beta = \frac km \cosh (mx), \qquad \e^\omega = \frac nk \sin (kt), \qquad
			k, m, n  > 0, 
\eeq
  where we have denoted $C_1 = m^2$, $C_2 = n^2$ (both $C_1$ and $C_2$ should be positive
  for \eqs \rf{int1} to be meaningful). The properties of this solution are to a large extent the same as 
  those of the branch $h < 0$ in Example 1, but a significant difference is the emergence of 
  $g_{tt} =  e^{2\beta(x)}$ that grows together with $r^2(x)$ at large $|x|$ somewhat similarly 
  to the static anti-de Sitter metric.

\medskip\noi
{\bf Example 3: $b\not\in \{0, 1/2, 1\}, C_1 = C_2 =0$.} In this case \eqs \rf{int1} are also easily 
  integrated in elementary functions, but the first equation implies $\beta'^2 >0$, that is,
  no throats are possible in the spacial sections of this space-time. Since we are seeking 
  \wh\ solutions, we do not consider this case any more.

\medskip
  It is clear that at the throat in all cases the gradient of the spherical radius $r(x,t)$ treated 
  as a scalar function in 2D space-time parametrized by $t$ and $x$ is timelike since $r'=0$ while 
  $\dot r \ne 0$. Thus at least a certain neighborhood of the throat is a so-called T-region, i.e., a 
  region where $r(x,t)$ may be chosen as a temporal coordinate (as happens, e.g., inside a 
  Schwarzschild horizon), and the geometry is thus cosmological in nature,
  resembling Kantowski-Sachs \sph\ cosmologies \cite{FN-BH, birk-16}. 
  Does it mean that a T-region covers the whole space-time? An answer apparently depends 
  on the parameters involved. For example, for the solution \rf{isotr} we have 
\beq
                    \d^\alpha r \d_\alpha r 
             = r_0^2 \cosh^2(kx) \biggl[\e^{2\omega(t)}\,h^2 \sign h + \frac {k^2}{\cosh^2(kx)}\biggr].
\eeq
  Thus in models with $h \geq 0$ we have $\d^\alpha r \d_\alpha r > 0$, i.e., the gradient of $r$ is 
  timelike in the whole space-time, it is a global T-region. On the contrary, in the case $h < 0$ 
  we have $\d^\alpha r \d_\alpha r = 0$ on the surfaces $x = \pm x_h(t)$ (apparent horizons)
  defined by the relation $ \sin^2 (ht) = 1/\cosh^2(kx) $, and there are R-regions at $|x| > |x_h(t)|$,
  in this sense the space-time is of \bh\ type.    
 
\subsection{Electromagnetic fields}

  Let us return to NED as the matter source of gravity. It is easy to notice that 
  the suitable form of $L(f)$ depends on the values of $q_e$ and $q_m$. 
  As follows from the expressions \rf{ff}, the magnetic field ${\bf B}$ is regular at all finite $r$,
  while the electric field ${\bf E}$ can tend to infinity not only where $r\to 0$ but also at zeros
  of the derivative $L_f$ if any. 

  Pure magnetic solutions are the simplest, just as in the static case. Indeed, assuming $q_e =0$ 
  and $q_m \ne 0$, we have $L(f) = 2R^\theta_\theta$, and using \rf{int1} and \rf{int2},
  it is easy to verify that $R^\theta_\theta$ is expressed in terms of $r = \e^{\beta+\omega}$ rather 
  than separately in terms of $\beta(x)$ and $\omega(t)$: 
\bearr                                  \label{R22a}
       R^\theta_\theta = r^{-2} + (2b{-}3) C_1 r^{-2b} + (2b{-}1)C_2 r^{2(b-1)},  \qquad b \ne 1/2,         
\yyy                                      \label{R22b}
       R^\theta_\theta = r^{-2} +  r^{-1} [2(C_4 - C_3) +  k^2 \ln (r/r_0)],         \qquad b = 1/2,         
\ear
  where $r_0$ is an arbitrary constant introduced for dimensional considerations, and its choice can 
  is related to the arbitrariness of the constant $C_4-C_3$.  
  Since $f = 2q_m^2/r^4$, it is straightforward to find $L(f)$ by substituting $r = (af)^{-1/4}$, 
  where $a= (2q_m^2)^{-1}$: 
\bearr                                  \label{Lm1}
       \Half L(f) = (af)^{1/2} + (2b{-}1)C_2 (af)^{(1-b)/2} + (2b{-}3) C_1 (af)^{b/2}, \qquad b \ne 1/2,         
\yyy                                      \label{Lm2}
       \Half L(f) = (af)^{1/2} + (af)^{1/4} [2 (C_4-C_3) +  k^2 \ln (f/f_0)],     \qquad b = 1/2,         
\ear
  where $f_0$ is introduced similarly to $r_0$. It is the full set of NED Lagrangians that lead to
   magnetic solutions of GR/NED equations with the metric \rf{ds3}. We see that none of 
  these $L(f)$ possess a Maxwell behavior at small $f$. 
 
  The corresponding electric solutions with the same metric pertain to other versions of NED,
  as follows from the FP duality described in Section 2. For the same metric \rf{ds3} we now have  
  $H = - 2R^\theta_\theta$, which, as we saw, is a function of $r$, and since now $p = -2q_e^2/r^4$,
  we know the function $H(p)$ which is the same as $-L(f)$ in \rf{Lm1} and \rf{Lm2}, where 
  $a$ should be replaced with $c = -1/(2q_e^2)$, and $f$ with $p$, so that
\bearr                                  \label{Hp1}
       -\Half H(p) = (cp)^{1/2} + (2b{-}1)C_2 (cp)^{(1-b)/2}+ (2b{-}3) C_1 (cp)^{b/2},\qquad b \ne 1/2,         
\yyy                                      \label{Hp2}
       -\Half H(p) = (cp)^{1/2} + (cp)^{1/4}[2 (C_4-C_3) +  k^2 \ln (p/p_0)],  \qquad b = 1/2,         
\ear
  Then, both $L$ and $f$ can in principle be found according to \rf{dual} as well as the function $f(p)$, 
  and, as in the static case, there should be as many different Lagrangians $L(f)$ in different parts of 
  space-time as is the number of monotonicity ranges of $f(p)$. From \rf{Hp1} and \rf{Hp2} we can 
  find $L$ in terms of $p$ (either as $L = 2p H_p -H$ or from the Einstein equations giving, for electric
  solutions, $L = 2R^t_t$), so that
\bearr                  \label{Lp1}
        \Half L  =  C_1(b-1)(3-2b) r^{-2b} + C_2 b(2b+1) r^{2(b-1)},   \qquad b \ne 1/2,         
\yyy			\label{Lp2}
	\Half L  = \frac 1r \Big[ -3 + C_4 - C_3 + 2k^2 \ln (r/r_0) \Big], \qquad b = 1/2.             
\ear
  and substituting $r = (cp)^{-1/4}$. The function $f(p)$ is also easily found since $f(p) = p H_p^2$, 
  and $H(p)$ is known. However, it is in general hard to find $L(f)$ and $p(f)$ in an explicit form 
  since finding the inverse of $f(p)$ requires solving a transcendental equation. 

  Let us look what happens in the above two simple examples, $b=1$ and $b=0$.   

\medskip\noi
{\bf Example 1: $b =1$,} with the metric \rf{ds4}, \rf{isotr}. For magnetic \whs, \eqn{Lm1}
  gives
\beq                 \label{L-ex1}
                      L(f) = 2(1-C_1) (af)^{1/2} + 2C_2,  
\eeq
  a Lagrangian of the form previously used in a number of studies, see, e.g., Refs. 
  \refcite{hendi-13}, \refcite{guen-14} and references therein. For magnetic wormholes supported 
  by NED in 2+1 dimensions such a Lagrangian was considered in Ref. \refcite{mazh-17}.

  For electric \whs\ with the same metric the function $-H(p)$ has the form \rf{L-ex1}  with 
  the substitution $af \mapsto cp$, see above. Assuming $C_1 \ne 1$, we obtain 
  $H_p \sim |p|^{-1/2} \sim r^2$, hence $ f = p H_p^2 =\const$.\footnote
	{According to \rf{ff}, surprisingly, the physically meaningful electric field strength ${\bf E}$ 
         is constant in this highly inhomogeneous and nonstatic space-time. It has been asserted 
         (e.g., in Ref. \refcite{Arel-06}) that the electric field is singular at the throat in this solution; 
         but this assertion applies to the parametrization-dependent quantity $F_{tr}$ in the particular
         coordinates used there rather than ${\bf E} = (F_{tr}F^{rt})^{1/2}$.}
  This strange result still does not immediately lead to a contradiction because by \eqn{Lp1} we have 
  also $L = 3C_2 = \const$. So this solution corresponds to a constant $f = F\mn F\MN$ at which 
  the function $L(f)$ has a certain particular value, but at other values of $f$ the function $L(f)$ 
  is not defined. However, since $L_f = 1/H_p\sim 1/r^2$, the derivative $L_f$ has different values 
  at different space-time points, and since $f$ is fixed, it is inconsistent with any well-defined function
  $L(f)$. This makes the Lagrangian formulation of the theory ill-defined for this electric solution.

\medskip\noi
{\bf Example 2: $b =0$,} with the metric \rf{ds5}, \rf{ds5a}. For magnetic \whs, \eqn{Lm1}
  gives
\beq                 \label{L-ex2}
                      L(f) = 2(1-C_2) (af)^{1/2} -3C_1,  
\eeq
  in full similarity with \rf{L-ex1}, up to the choice of the integration constants $C_1, C_2$. So this 
  is one more kind of magnetic \wh\ geometry supported by NED with such Lagrangians. As to
  electric \whs\ with the same geometry, the function $L(f)$ for them is again ill-defined for the 
  same reasons as in Example 1.

  For dyonic solutions with the same metrics, the problem of finding suitable NED formulations is 
  more difficult and will not be considered here.
   
\section{Concluding remarks}

  1. We have recalled the problem of obtaining regular \ssph\ black hole solutions in GR with NED 
  as a source of gravity and stressed the fundamental distinction between electric and magnetic 
  fields, connected with the absence of the familiar duality inherent to the Maxwell electrodynamics.
  In the latter, knowing, for example, a magnetic field, it is straightforward to obtain any mixture of 
  electric and magnetic fields by duality rotations. Unlike that, in the NED framework, FP duality 
  connects only pure electric and pure magnetic configurations with the same metric but sourced 
  by different versions of NED. It also happens that only pure magnetic solutions lead to completely 
  regular configurations flat at infinity since their electric counterparts exhibit undesired features
  at some intermediate radii. Next, it turns out that finding dyonic configurations in NED/GR is  
  quite a nontrivial problem, and only some special examples of such solutions are known.

  2. In our search for evolving \wh\ solutions, we have obtained a new family of geometries, which
  are in general not conformally static, except for a special case where they reduce to the known 
  solutions of Ref. \refcite{Arel-06}. Under the metric ansatz \rf{ds3}, these geometries represent the
  general solution of GR for Dymnikova's vacuum, defined by its transformation properties in the 
  ($x^0, x^1$) 2D subspace, similar to the properties of the cosmological constant in the whole 4D 
  space-time.\footnote
	{In a sense, it is half of the whole set of solutions because we chose $k^2 >0$ as the
	  separation constant in \eqn{bom} in order to obtain a minimum of $\e^{\beta(x)}$. 
           Solutions with the other sign of the separation constant certainly exist as well.}
  A general feature of these geometries is the existence of a throat in their spatial sections. 
  However, their time evolution in general contains cosmological-type singularities. 
  It has been shown that they can neither have a bounce of $r(x,t)$ nor begin or end with its finite 
  value as $t \to -\infty$. It is not excluded that they have a ``remote singularity'', that is, a zero 
  value of $r$ as $t \to \pm \infty$, but even this opportunity may be ruled out in a future study.
  The existence of singularities at finite times is confirmed by examples given by \eqs \rf{isotr}
  and \rf{ds5a}.  	 

  3. All that was independent of the particular choice of Dymnikova's vacuum. If we apply NED for its
  implementation, it is rather easy to obtain these solutions with pure magnetic fields, but the choice of
  suitable NED Lagrangians is quite narrow and is expressed in \eqs \rf{Lm1} and \rf{Lm2}. None 
  of these Lagrangians have a Maxwell weak field limit.
 
  4. We have extended the FP duality, previously formulated for static systems, to time-dependent 
  ones and used it for comparison between the electric and magnetic evolving \wh\ solutions.
  The electric solutions are obtained in the ``Hamiltonian'' formulation of NED in the same way as
  magnetic ones in its Lagrangian formulation. However, it is in general hard to obtain a Lagrangian
  $L(f)$ for a given electric solution due to a necessity to deal with transcendental equations.
  Moreover, in particular examples the Lagrangian formulation is ill-defined for electric solutions.

  To conclude, the NED/GR system exhibits some unusual and unexpected features and deserves 
  a further study and discussion.   
 
\subsection*{Acknowledgments}

  I thank Milena Skvortsova, Sergei Bolokhov and Sergei Rubin for helpful discussions. 
  The work was partly performed within the framework of the Center FRPP 
  supported by MEPhI Academic Excellence Project (contract No. 02.a03.21.0005, 27.08.2013).
  The work was also funded by the RUDN University Program 5-100 and by RFBR grant 16-02-00602.


\end{document}